\documentclass[a4paper,12pt]{article}
\usepackage{graphicx}

\def\FRAC#1#2{\leavevmode\kern-.1em
 \raise.5ex\hbox{\the\scriptfont0 #1}\kern-.1em
  /\kern-.15em\lower.32ex\hbox{\the\scriptfont0 #2}}
\begin{document}

\title
{\bf Trajectory of the harmonic oscillator in the Schr\"odinger
wave
}

\author
{
Yoshio {\sc Nishiyama}\\
Department of Science Education,
 Faculty of Education and Human Sciences\\
 Yokohama National University, Yokohama,
  240-8501 Japan
\thanks{e-mail: nisiyama@ed.ynu.ac.jp}
}

\sloppy
\maketitle

\renewcommand{\abstractname}{}
\begin{abstract}
 A trajectory of a harmonic oscillator obeying the Schr\"odinger wave
equation is exactly derived and illustrated.
 The trajectory resembles well the classical orbit between the turning
points,  and also runs through the tunneling region.
 The dynamics of the `particle' motion and the wave function associated
with the motion  are proposed.
 The period of a round trip on the trajectory is exactly equal to that
obtained in classical mechanics.
\end{abstract}

\section{Introduction}
In general the motion of a particle whose beam shows an interference
 phenomenon is described by the wave equation.
On the other hand the orbit of the particle described in classical
 mechanics plays an important role in the fields such as electron
 optics, particle accelerators, radiation from electron beams and
 so forth.
 The significance of the scattering wave function, wave-optical approach,
 was clarified by comparison with  the classical wave function,
 ray-optical approach, by Gordon.~\cite{GORDON}

 Geometrical optics  describes the movement of the light corpuscle
  in the space as the ray of light.
 Formulas for the ray of light derive  from the eikonal equation.
  This corresponds to the Hamilton characteristic function
  for the particle motion,~\cite{BornW3}
  which can be derived  from the Schr\"odinger wave function in the WKB
  approximation for the system with the stationary potential.

  The concept of the orbit or the ray of light in the wave phenomena
  gives a comprehensive image and physical insight of the process,
  although it has been an approximate
  idea.~\cite{Brekhovskikh80,KravOrlo80,Kravtsov88}
  The field ion microscope and scanning tunneling microscope are now
   available to image an atomic movement
    on solid surfaces.~\cite{Tsong,Ganz}
   An atom can be picked up and moved to an arbitrary place.~\cite{Eigler}
  These suggest the possibility of the precise description of the
  motion of the particle even in the atomic scale.

  A proposal of extension of the light ray to the shadow region had
 been made in order to clarify the physics of the
 diffraction.~\cite{Keller}
  The relation between the wave and the light ray or orbit should be
 investigated  more carefully.
   Trajectory in the optical wave, extension of a ray of light, was developed
     by generalizing the eikonal to the mode characteristic function
      with dynamical assumptions.~\cite{Nishiyama}

 If the motion of the particle is restricted to that in the classical
  region where classical mechanics is the case, the orbital motion
  is derived from the Hamilton  characteristic function
  by the Hamilton-Jacobi theory for the system with the stationary
  potential.~\cite{Goldstein}
 If the characteristic function could be generalized to that valid
 for every space region, it would be able to pinpoint the motion
   of the particle even in the tunneling region.~\cite{YN6365}

 Along the similar way of thinking, Bohm proposed a quantum theory with
 ``hidden variables'' to suggest objective description of individual
  systems at a quantum level of accuracy.~\cite{Bohm}
It has made a general scheme of the causal interpretation but
 not given a concrete trajectory with the dynamical behavior of
  a `particle' in the space-time region from the wave equation.
The goal of every causal theory or interpretation as to the quantum
 mechanics is a complete description of an individual real situation
as it exists independently of acts of observation.~\cite{Holland}
The motivation is summarized in the Einstein's feeling that
the statistical prediction of the quantum theory is correct but
by supplying the missing elements, it could be in principle got
beyond statistics to a determinate theory.~\cite{Bohm85}
To fulfil the aim it might be necessary to exploit the suitable
mathematical tools to describe the particle motion in quantum
mechanics.

 In the present paper, a trajectory of the harmonic oscillator in
 one and two dimension is exactly derived from the Schr\"odinger
 wave function.
 The trajectory and dynamical motion are compared with the classical
 orbit and dynamical behavior in the classical region.
The relation between the traveling waves associated with the `particle'
 motion and usual stationary wave function is discussed.

\section{Dynamics and wave function}\label{sec:QDWF}
The way to derive the trajectory from the wave equation is described
 and  the significance of the wave function is discussed.
The Schr\"odinger wave equation for a particle in a stationary potential,
 $V$, in one spatial dimension, is
\begin{eqnarray}
  i\hbar {\partial \over \partial t} \Psi = \left(
   - {\hbar^2 \over 2m}\frac{\partial^2}{\partial x^2} + V \right) \Psi
   \equiv H \Psi.
\end{eqnarray}
The equation is separable in $t$ and $ x$.
 A wave function for a `stationary' state is written as
\begin{eqnarray}
 \Psi (x, t) &=& \exp(-iEt/\hbar) \Phi(x, E),
\end{eqnarray}
 where  $E$ is a constant of separation.
 Function $\Phi$ is a general solution of an ordinary differential
 equation of second order.
 The boundary condition that the function be bounded and continuous
 everywhere in the domain defined is not imposed on the function yet.
 Function $\Phi$ is the `eigenfunction' of the Hamiltonian operator $H$.
  Constant $E$ is the `eigenvalue' and the energy of the `stationary'
   state.
 The wave function $\Psi$ determines the `eigenstate', or the `mode',
  specified by $E$.
 The constant $E$ should be called the mode parameter.

 If the parameter takes specific, discrete, values,
  the wave function satisfies the boundary condition.
  The state with a specific parameter is usually called a
 stationary state or the mode, and function $\Phi$ is the eigenfunction.
  For the present, the parameter is assumed to be a real number and the
   words, such as stationary, eigen, mode and particle are used
   with quotation marks.

 The wave function of the form of
\begin{eqnarray}
   \Phi(x, E) = \sqrt{\rho (x, E)} \exp [i\,W (x, E)] \qquad
                                    \label{eq:Wjxj}
\end{eqnarray}
 is sought, where functions $\rho $ and $W$ are real number.
 Function $W$ satisfies approximately
 \begin{eqnarray}
   E = {1 \over 2m}\left(\hbar \frac{\partial W}{\partial x} \right)^2
     + V   \label{eq:HJacobi}
  \end{eqnarray}
  in the WKB approximation.
 The region where this equation holds
 should be called the classical region.
  This is the Hamilton-Jacobi equation with energy $E$.~\cite{Goldstein}
  If the  solution for $\hbar W$ of this equation is written
   as $W_{\rm cl}$,  the Hamilton characteristic function is
    given by~\cite{Goldstein}
 \begin{eqnarray}
 W_{\rm cl}(x, E) &=& \int^x \sqrt{2 m (E - V)} dx.
                            \label{eq:WHcf}
 \end{eqnarray}
 Function $W$ satisfies the nonlinear equation (4) in the reference of
 Bohm~\cite{Bohm}, expressed in the form $S = -Et + \hbar W$,
  the Hamilton-Jacobi-like equation with the ``quantum-mechanical''
   potential.
  Function $W$ is assumed to satisfy
     the condition that in the classical region of $x$
 \begin{eqnarray}
    \hbar W(x, E) \simeq  W(x, E)_{\rm cl},
                                \label{eq:WWcl}
  \end{eqnarray}
 except an additional constant independent of $x$ and the mode
  parameter.
 If function $W$ can be determined uniquely, function $W$ should
 be named the mode characteristic function (mcf) for the system.
 The mcf $W(x)$ is derived from the wave function $\Phi(x)$ as follows.
 Since the wave equation  is the ordinary
 differential equation of 2nd order, the function $\Phi$ is composed of two
 linearly independent solutions, say $u_1$ and $u_2$,
\begin{eqnarray}
 \Phi(x, E) = a\, u_1(x, E) + b\, u_2(x, E),           \label{eq:LIwvfn}
\end{eqnarray}
 where $a$ and $b$ are complex constants.
  Constants $a$ and $b$ should be determined by the following assumptions
 from the theoretical point of view.

Assumption 1: Function $\hbar W$ in the classical region must be
 as approximate to the corresponding characteristic function
  $W_{\rm cl}$ as possible in the sense of relation (\ref{eq:WWcl}).

Assumption 2: The results derived from the equations of motion
 for $W$ defined by Eq.$\!$ (\ref{eq:MotionS}) should be as
  approximate to the ones in classical mechanics as possible.

 Function $W(x)$ should be determined as the phase or argument
   of function $\Phi$ like Eq.(\ref{eq:Wjxj}).
  The function $\Psi$ consisting of this $\Phi $ may be said
 to represent the traveling wave associated with the motion of
 a `particle' in the `mode' $E$.

The equation of motion for the `particle' is assumed
 \begin{eqnarray}
 {\partial W \over \partial E} = {1 \over \hbar}(t - t_0),
                  \label{eq:MotionS}
 \end{eqnarray}
where $ t_0$ is a constant (independent of $t$)
 that should be determined by the initial condition for the system.
 Variable $t$ should be the dynamical time for the system.

 Dynamical time $t$ should be determined so as to increase monotonically
 as the `particle' moves.
 From equation (\ref{eq:MotionS}), a phase velocity
 \begin{eqnarray}
 {\partial x \over \partial t} = \left( \hbar
    {\partial^2 W(x, E) \over
                 \partial x \partial E} \right)^{-1},
 \end{eqnarray}
   is obtained.
If the `particle' starts from a position, say  $x_{0}$,
  at an initial time and $\partial t /\partial x \ge 0 $
   for $x \le x_{b} $,
 it could be considered that the `particle' runs from $x_{0}$ to $x_{b}$.
Here $x_{b}$ stands for a turning point or an endpoint of the potential.
After reaching $x_{b}$, the `particle' returns to $x_{0}$
 with the mcf of
 $- W(x, E) + 2 W(x_{b}, E)$,
  which guarantees the monotone increase of time $t$ given by
   equation (\ref{eq:MotionS}).

The wave function associated with the `particle' motion in a bound state
 should be  described as follows.
If the function $\Phi$ (\ref{eq:Wjxj}) represents the motion
 of the `particle' traveling to the right, function
 $\Phi^* = \sqrt{\rho} \exp [- i\,W]$,
  which is also the solution of the wave equation,
 stands for the motion to the reverse direction.
Let the endpoints of the potential in the $x$ coordinate be $a$ and
 $b$.
 Then the `particle' moves in the region $a \le x \le b$.
Let it start from a point $x_{0}$ to the positive $x$ direction and
the mcf be $W(x)$.
The traveling wave associated with the returning motion from $b$
to $x_{0}$ or $a$ should be given by
 $\sqrt{\rho(x, E)} \exp ( i\,[ -W(x, E) + 2W(b, E) ] )$.

The wave function observed at $x (\ge x_{0})$ is assumed to be
 the superposition of the traveling waves of either motion~\cite{GORDON}
 \begin{eqnarray}
     \Phi(x, E) - \Phi^*(x, E)\exp[ i\,2W(b, E) ].
                    \label{eq:xjbj}
\end{eqnarray}
 This is finite, zero, at $b$.
  If the `particle' turns at $a$ and runs to $x_{0}$ or $b$,
 the mcf should be given by $W(x, E) - 2W(a, E) + 2W(b, E)$.
  The wave function for $x (\le x_{0})$ is thus to be written as
 \begin{eqnarray}
     - \Phi^*(x, E)\exp[ i\,2W(b, E) ]
     + \Phi(x, E) \exp \left( i\, [- 2W(a, E) + 2W(b, E)] \right).
                            \label{eq:xjaj}
\end{eqnarray}
 This is finite, zero, at $a$.
 The wave functions (\ref{eq:xjbj}) and (\ref{eq:xjaj}) are bounded
  for $a \le x \le b$ for any `mode',
 although they might not be continuous at $x_{0}$.

 For $x\, ( \ge x_{0})$ the traveling wave associated with a round
 trip of the `particle' gets the shift in phase by
  $ - 2W(a, E) + 2W(b, E) $.
 The wave function associated with $N$ round trips of the `particle' motion
  would result in, being averaged for one cycle,
 \begin{eqnarray}
  \lefteqn{ {1 \over N}\,
  \frac{ 1 - \exp \left( i\, N [- 2W(a, E) + 2W(b, E)] \right)}
       { 1 - \exp \left( i\, [- 2W(a, E) + 2W(b, E) ] \right)} }
                \nonumber\\  & &{}\times
    \left( \Phi(x, E) - \Phi^*(x, E)\exp[i\, 2W(b, E)] \right).
                                \label{eq:resona}
 \end{eqnarray}
  If the number $N$ is large, this shows a sharp resonance if $E$
 satisfies
 \begin{eqnarray}
   - 2W(a, E) + 2W(b, E) = 2\pi \times\, {\rm integer}.
                                \label{eq:resonance}
 \end{eqnarray}
  This resonance condition gives rise to the stationary state
 and the eigenfunction.
 This is the exact version of the Bohr-Sommerfeld quantum condition.
 An approximate but a little general expression for it had been
 presented  previously.\cite{KelRubi}
 It  might as well be interpreted that the observed wave should be
  proportional to the wave function mentioned above.

\section{ Simple harmonic oscillator}\label{sec:SHO}
 The  trajectory of the simple harmonic oscillator is discussed.
  The wave equation for the oscillator with mass $m$ for energy $E$
   is written as
\begin{eqnarray}
    \left ( - {\hbar^2 \over 2m} {\partial ^2 \over \partial x^2}
    + {m \over 2}\omega ^2 x^2 \right ) \Phi(x) = E \Phi(x).
                                \label{eq:SHOeq}
\end{eqnarray}
A solution of this equation of the form (\ref{eq:LIwvfn}) that leads to
 function (\ref{eq:Wjxj}) is
\begin{eqnarray}
  \Phi^{(+)}(x) = \left [ M(a, \FRAC{1}{2}, z) - i {\Gamma(1-a) \over
       \sqrt{\pi} \Gamma(\FRAC{1}{2}-a)} V(a, \FRAC{1}{2}, x) \right ]
                    \exp(-z/2),
                           \label{eq:SHORGwf}
\end{eqnarray}
where $\Gamma(x)$ is the gamma function and
\begin{eqnarray}
  a =  \FRAC{1}{4} \left(1 - 2E / \hbar  \omega \right), \quad
   z = m \omega x^2 / \hbar.
\end{eqnarray}
Functions $M$ and $V$ are linearly independent confluent hypergeometric
 functions.~\cite{AbraSteg}
 The former is the Kummer function, which is expressed in the series form
\begin{eqnarray}
    M(a, \FRAC{1}{2}, z) = \sum_{k=0}^{\infty}
       {(a)_k z^k \over ( \FRAC{1}{2})_k k! }
\end{eqnarray}
and the latter is defined as~\cite{Nishiyama}
\begin{eqnarray}
 V(a, \FRAC{1}{2}, x) = \Gamma (- \FRAC{1}{2}) \sqrt{m\omega /\hbar}\, x
        M(\FRAC{1}{2} + a, \FRAC{3}{2}, z).
\end{eqnarray}
Expression (\ref{eq:SHORGwf}) is the `eigenfunction' with
 `eigenvalue' $E$, not always bounded at $x = \infty$, of
  Eq.$\!$ (\ref{eq:SHOeq}).

 The mcf $W(x,E)$ satisfying two assumptions mentioned in section
 \ref{sec:QDWF} is given by
\begin{eqnarray}
  W(x,E) &=&  \arctan \left [ {- \Gamma (1-a) \over
  \sqrt{\pi} \Gamma ( \FRAC{1}{2} - a)}
   {V(a, \FRAC{1}{2}, x) \over M(a, \FRAC{1}{2}, z)} \right ]
                                                        \nonumber \\
   & & \equiv \arctan [F(a, \FRAC{1}{2}, x)]   \label{eq:mcfSHO}
\end{eqnarray}
 This mcf multiplied by $\hbar$ is well approximate to $W_{\rm cl}(x,E)$
  with $W_{\rm cl}(0,E) = 0$ in the classical region or
 in the neighborhood of 
  $x = 0$.
The validity of the mcf will be recognized in the following discussion.

The equation of motion is given by Eq.$\!$ (\ref{eq:MotionS}), or
\begin{eqnarray}
  t &=& {1 \over \omega} \Biggl[ \Psi (1-a) - \Psi( \FRAC{1}{2} - a)
  - {\partial \log M(\FRAC{1}{2} + a, \FRAC{3}{2}, z) \over \partial a }
  + {\partial \log M(a, \FRAC{1}{2}, z) \over \partial a} \Biggr]
                                                         \nonumber \\
   & & \times {\Gamma (1-a) \over \Gamma (\FRAC{1}{2} - a)}
    {M(\FRAC{1}{2} + a, \FRAC{3}{2}, z) \over M(a, \FRAC{1}{2}, z)
   [ 1 + F(a, \FRAC{1}{2}, x)^2 ]},
                             \label{eq:tvsx}
\end{eqnarray}
 where $\Psi (x) = d\log \Gamma (x)/dx$ is the psi function.~\cite{AbraSteg}
 The oscillator has been assumed to be at $x = 0$ at $t = 0$.

By using the asymptotic form of the confluent hypergeometric
  functions,~\cite{AbraSteg} it is obtained for $x$ large
\begin{eqnarray}
 F(a, \FRAC{1}{2}, x) \approx \tan \left[{\pi \over 4} \left(1
  + {2E \over \hbar \omega } \right) \right].  \label{eq:AsympF}
\end{eqnarray}
It is thus obtained that $t(x = \infty) = \pi /2\omega $.
The $x$ dependence of function $t(x)$ is monotone everywhere, as proved
 by a computer calculation.
 Therefore it can be considered that the oscillator moves between the end
 points, $x = -\infty$ and $\infty$, without interruption.

 If the oscillator starts from $x = 0$ at $t = 0$ and goes to $x = \infty$,
 it returns there and comes back to $x = -\infty$.
  Then it goes back to $x = 0$.
  The mcf $W(x,E)$ for one cycle is written as follows:
\begin{eqnarray}
 W(x,E) = \left\{
       \begin{array}{@{\,}ll}
            \arctan [F(a, \FRAC{1}{2}, x)]  & \mbox{ ($x = 0 \to \infty$),} \\
         2W(\infty,E) - \arctan [F(a, \FRAC{1}{2}, x)]
          & \mbox{ ($x = \infty \to  -\infty $), } \\
         4W(\infty ,E) + \arctan [F(a, \FRAC{1}{2}, x)]
          & \mbox{ ($x = -\infty \to x = 0$). }
        \end{array}
        \right.                 \label{eq:SHOx}
\end{eqnarray}
These are determined so that time $t$ increases as the oscillator runs
 along the trajectory.
 From the asymptotic form of  $F(a, \FRAC{1}{2}, x)$ for $x$ large,
 (\ref{eq:AsympF}),  and $F(a, \FRAC{1}{2}, 0) = 0$, it is seen
 that the period of the motion is equal to $2\pi /\omega $
 that is just equal to that in  classical mechanics.
  The oscillator runs throughout the space between $x = -\infty$ and $\infty$
  and the phase velocity $dx/dt$ in the tunneling region is very large.

The variation of the mcf after one cycle stands for the change
 of the phase  of the wave function.
It is at any point
\begin{eqnarray}
 -\,2 W(-\infty ,E) + 2W(\infty ,E) =
 4W(\infty ,E) = \pi \left( 1 + 2E / \hbar \omega  \right).
\end{eqnarray}
The resonance condition of the wave function (\ref{eq:resonance})
 leads to the eigenvalues of energy.
 Thus it holds for the stationary state
\begin{eqnarray}
   E = \hbar \omega \left( n + \FRAC{1}{2} \right),
   \qquad  n = 0,1,2, \cdots.
\end{eqnarray}

\begin{figure}[htbp]
\begin{center}
  \includegraphics[width=6.5cm,height=5cm]{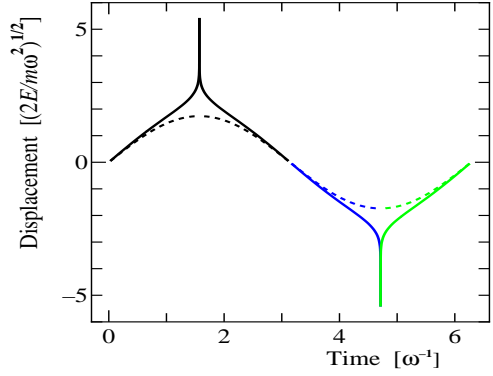}    \\
\end{center}
\caption{ Position versus time of a harmonic oscillator with
  $E = \FRAC{3}{2} \hbar \omega$. Solid line stands for the trajectory
  and broken line for the corresponding classical orbit. }
\label{fig:Hoscxvst}
\end{figure}

If equation (\ref{eq:tvsx}) 
 is solved inversely for $x$ as a function of $t$,
 the function $x(t)$ is well approximated by the classical oscillation
\begin{eqnarray}
   x = \sqrt{ 2E / m \omega^2} \sin \omega t.
                      \label{eq:xvstcl}
\end{eqnarray}
Expressions (\ref{eq:tvsx}) and (\ref{eq:xvstcl}) are illustrated
 in Fig.~\ref{fig:Hoscxvst}.
The discrepancy between the two occurs at about times,
 $t = \pi/2\omega \times$ half integer, when the oscillator
  is running through the tunneling region.

 The similarity suggests that the mcf (\ref{eq:mcfSHO}), satisfying
  assumptions in section \ref{sec:QDWF}, is the correct one
 for the  harmonic oscillator.

\section{Wave function}\label{sec:WaveFn}
The wave function (\ref{eq:SHORGwf}) multiplied by $\exp(-\,i\,Et/\hbar)$
 describes a wave traveling to the right.
According to the mcf (\ref{eq:SHOx}) it may be considered that
 it represents the wave associated with the oscillator motion for $x=0$ to $\infty$.

 A wave traveling to the left associated with the oscillator motion
  from $x = \infty$ to $-\infty$ should be given by
\begin{eqnarray}
 \Phi^{(-)}(x, E)
   &=& \left [ M(a, \FRAC{1}{2}, z) + i\, \frac{\Gamma(1-a)}{
     \sqrt{\pi} \Gamma(\FRAC{1}{2} - a)} V(a, \FRAC{1}{2}, x) \right ]
                               \nonumber \\
   & &{}\times \exp \left[ -z/2 + i\, 2W(\infty, E) \right].
                           \label{eq:SHOLGwf}
\end{eqnarray}
 A linear combination with $\Phi^{(+)}$ (\ref{eq:SHORGwf})
like expression (\ref{eq:xjbj}),
\begin{eqnarray}
 \Phi_1(x, E) &=& \Phi^{(+)}(x, E) - \Phi^{(-)}(x, E)   \nonumber\\
              &=& 2 \exp(- i \pi a - \FRAC{1}{2} z) \Bigl [
             M(a, \FRAC{1}{2}, z) \cos \pi a            \nonumber\\
         & &{} + {\Gamma(1-a) \over \sqrt{\pi} \Gamma(\FRAC{1}{2} - a)}
                    V(a, \FRAC{1}{2}, x)\sin \pi a  \Bigr ],
                               \label{eq:SHOwfR}
\end{eqnarray}
 gives rise to  a wave function finite for $0 \le x \le \infty$.
The boundedness at $x = \infty$ is seen as follows.
Expression (\ref{eq:SHOwfR}) can be rewritten as
\begin{eqnarray}
     \Phi_1(x, E)
     = {2 \sqrt{\pi} \over \Gamma( \FRAC{1}{2} - a)}
   \exp(- i\, \pi a - z / 2) x U(a + \FRAC{1}{2}, \FRAC{3}{2}, z).
\end{eqnarray}
Function $U$ is the Kummer function.~\cite{AbraSteg}
 From the asymptotic form of function $U$  for $x$ large~\cite{AbraSteg}
 it is found that
\begin{eqnarray}
     \Phi_1(x, E) \simeq
    {2 \sqrt{\pi} \over \Gamma(\FRAC{1}{2} - a)}
     z^{-a} \exp (-i\, \pi a - \FRAC{1}{2} z),
\end{eqnarray}
 which tends to zero as $x$ tends to infinity.

\begin{figure}[htbp]
\begin{center}
  \includegraphics[width=7cm,height=5.5cm]{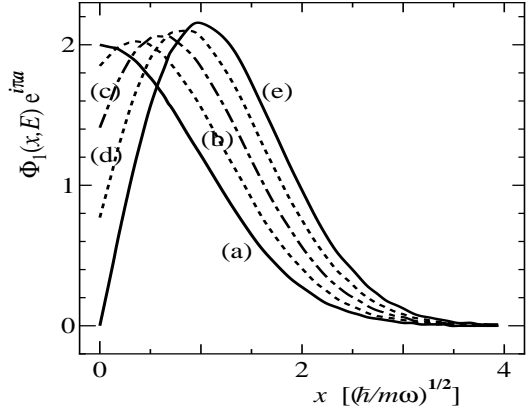} \\
\end{center}
\caption{ Wave function $\Phi_1(x, E)e^{i \pi a}$ vs $x$ for
 $E = \FRAC{1}{2}\hbar \omega \times $
 (a) 1,\, (b)1.5,\, (c) 2,\, (d) 2.5,\, (e) 3. }
\label{fig:wavef}
\end{figure}

For $x$ negative, an associated wave for the oscillator motion reflected
 at $x = -\infty$ and going back to $x = 0$ should be written as
\begin{eqnarray}
   \lefteqn{ \Phi^{(+)}(x, E)' }                      \nonumber \\
   &=& \left[ M(a, \FRAC{1}{2}, z) - i {\Gamma(1-a) \over
         \sqrt{\pi} \Gamma(\FRAC{1}{2} - a)} V(a, \FRAC{1}{2}, x) \right]
                   \exp[- z / 2 + i\, 4 W(\infty, E)].
\end{eqnarray}
 The wave function for $x$ negative is
\begin{eqnarray}
\lefteqn{\Phi_2(x, E) = \Phi^{(+)}(x, E)' - \Phi^{(-)}(x, E)}
        \nonumber\\ &=&
    2 \exp(- i 3\pi a - \FRAC{1}{2} z) \Bigl[
    M(a, \FRAC{1}{2}, z) \cos \pi a             \nonumber\\
    & &{} + {\Gamma(1-a) \over \sqrt{\pi} \Gamma(\FRAC{1}{2} - a)}
         V(a, \FRAC{1}{2}, x)\sin \pi a  \Bigr ].
       \label{eq:SHOwfL}
\end{eqnarray}
 Since $M(a, \FRAC{1}{2}, z)$ is an even function of $x$ and
  $ V(a, \FRAC{1}{2}, x)$ is an odd, there is a symmetry between
 functions $\Phi_1$ and $\Phi_2$
\begin{eqnarray}
   \Phi_2(x, E) = \Phi_1(-x, E) \exp( -i\, 2\pi a).
\end{eqnarray}
A wave function $\Phi_1$ except a constant but $E$-dependent factor
 is shown in Fig.~\ref{fig:wavef} for several $E$'s.

It can be seen from the figure that function $\Phi_1$ is smoothly
continuous to function $\Phi_2$ at $x = 0$ if
 $E = \FRAC{1}{2}\hbar \omega \times$ $[1$ or $3]$, which is
 the eigenstate.

 If $4 W(\infty, E)$ is a multiple of $2\pi $, in general,
 expression (\ref{eq:SHOwfR})
 becomes smoothly continuous to equation (\ref{eq:SHOwfL}) at $x=0$.
 The two expressions constitute the eigenfunction for the stationary state
 finite and smoothly continuous  for any space point with the eigenvalue
 $E$ discrete.

\section{ Two dimensional motion}\label{sec:2DHOS}
The simple harmonic oscillator runs only between
the end points, $x = - \infty$  and $x = \infty$.
 Except that it runs through tunneling regions, it oscillates
   like a classical particle.
 Here,  the harmonic oscillator moving in the space of two dimension
 is studied.
The extension is straightforward if the partial differential equation
 is separable in variables.

 The wave equation for the harmonic oscillator in two dimensional space
 with energy $E$ is given by 
\begin{eqnarray}
 E \Phi(x,y) = \left [ -\, {\hbar ^2 \over 2m} \left( {\partial ^2 \over
  \partial x^2} + {\partial ^2 \over \partial y^2} \right)
 + {m \over 2}(\omega_1^2 x^2 + \omega_2^2 y^2 ) \right ] \Phi(x,y),
\end{eqnarray}
 where $m$ is the mass, and $\omega_1$ and $\omega_2 $ are proper
 frequencies.
By introducing a constant of separation of variables, $E_2$,
the above equation can be decomposed into two equations
\begin{eqnarray}
 E_1\Phi_1(x,E_1) &=& \left( -\, {\hbar ^2 \over 2m}
                {\partial ^2 \over \partial x^2}
               + {m \over 2}\omega_1^2 x^2 \right) \Phi_1(x,E_1),    \\
 E_2\Phi_2(y,E_2) &=& \left( -\, {\hbar ^2 \over 2m}
                  {\partial ^2 \over \partial y^2}
                + {m \over 2}\omega _2^2 y^2 \right) \Phi_2(y,E_2),
\end{eqnarray}
where $E_1 = E -  E_2$.

By following the discussion of section \ref{sec:SHO},
the mcf in the `mode' $\,(E, E_2)$,
 is obtained as
\begin{eqnarray}
 W(x, y, E, E_2) = W_1(x, E, E_2) + W_2(y, E_2),
\end{eqnarray}
 where
\begin{eqnarray}
 W_1(x, E, E_2) = \arctan F \left( \FRAC{1}{4}(1 - 2E_1 / \hbar \omega_1 ),
        \FRAC{1}{2}, x \right),                \\
 W_2(y,E_2) = \arctan F \left( \FRAC{1}{4} (1 - 2E_2 / \hbar \omega_2),
       \FRAC{1}{2}, y \right).
\end{eqnarray}

The equations of motion are given by equation (\ref{eq:MotionS}), or
\begin{eqnarray}
 t = t_0 + \hbar {\partial W_1(x, E, E_2) \over \partial E}
    = t_2 + \hbar {\partial W_2(y,E_2) \over \partial E_2},
\end{eqnarray}
where $t_0$ and $t_2$ are constants to be determined by the initial
 condition.
The whole trajectory for $-\infty \le x \le \infty $ and
$-\infty \le y \le \infty $ can be obtained by using mcf's (\ref{eq:SHOx})
 for $x$ and $y$ coordinates.
 The projected motion of the `particle' onto the $x$ or $y$ coordinate
  is the periodic one with period of $2\pi /\omega_1$ or $2\pi /\omega_2$,
 respectively.
 If the ratio $\omega_1/\omega_2$ is irrational, the trajectory
   in the two dimensional space is not closed as is the case
     in classical mechanics.

\begin{figure}[t]
\begin{center}
  \includegraphics[width=6.5cm,height=5.5cm]{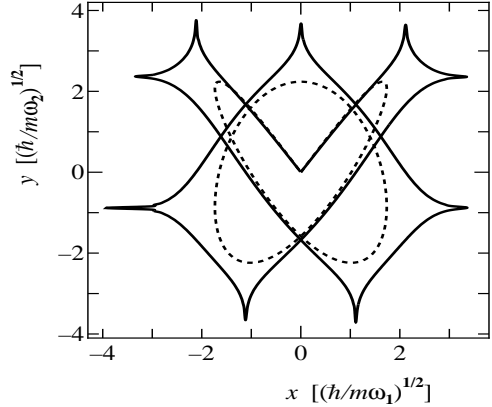}    \\
\end{center}
\caption{ The trajectory of a two-dimensional harmonic oscillator
with $E_1 = \FRAC{3}{2} \hbar \omega_1$, $E_2 = \FRAC{5}{2} \hbar\omega_2$
 and  $\omega_1/\omega_2 = 0.8$ \, (solid line), and the corresponding
 classical orbit (broken line).}
\label{fig:HoscDec}
\end{figure}

  An example of the trajectory for the system with parameters
  $E_1 = \FRAC{3}{2} \hbar \omega_1$ and $E_2 = 2 \hbar \omega_2$
  and $\omega_1/\omega_2 = 0.8$ is shown in Fig.~\ref{fig:HoscDec}.
   The oscillator is set to start from the origin at $t = 0$, or
  $t_0 = t_2 = 0$.
  In the figure the corresponding classical orbit with the same
    parameters is also drawn.
 It could be seen that the bigger the values of the parameters,
    the closer the trajectory and the classical orbit
     with each other, which shows the correspondence principle.

\section{Conclusion }\label{sec:Conclusion}
By generalizing the argument on optical wave,\cite{Nishiyama}
 the dynamics that should figure out a ray of `particle',
 a trajectory, in a `mode' of the Schr\"odinger wave equation
 of the completely separable form, especially in one spatial dimension,
 has been proposed.
The dynamics should be determined by the mcf and dynamical assumptions
 on it.
 The trajectory thus determined resembles well the orbit of
 the corresponding state in classical mechanics in the classical region.
 It runs also through a tunneling region.
 This is verified for the harmonic oscillator system.
 The period of a `particle' in the oscillator is exactly equal
 to that in classical mechanics.

The wave function bounded everywhere but not always continuous
 for any bound state could be made by superposing the traveling waves
   associated with the `particle' motion.
 The eigenfunction of the stationary state should be interpreted to be
  the wave function of the resonating state in the potential of the system.
 It is to be noted that all solutions of the wave equation are
  necessary in order to derive the mcf to get the trajectory.
 This suggests the role of all the solutions of the wave equation.
  Since the trajectory is determined by the phase or argument
  of the traveling wave function,
  it does not always show all the path of the energy transfer,
  which might be necessary
   for the real particle motion in the wave equation.

 The accordance of these characteristics
  between classical and quantum mechanics
   suggests the validity of the dynamics defined here.
 The main difference from the Bohm's theory~\cite{Bohm} stems
 from the dynamical assumption (\ref{eq:MotionS}).
 The significance and the consistency with the quantum theory
 should be verified for a more important system such as the one
 with the Coulomb potential.
 By treating the scattering problem, the statistical but an in principle
  determinate nature in quantum theory will be shown on getting
  the cross section.

It suggests the consistent existence of the trajectory in wave mechanics
 and  the significance of the relation between the particle motion
 and the wave function,
  although there may remain a lot to be considered about the observation
   process of particle and wave phenomena.\\

\noindent
{\bf Acknowledgments}\\
The author would like to express his sincere thanks
 to Drs. Y. Takano, S. Nakamura and T. Okabayashi
  for their continual encouragement during the long course of the work.


\newpage
\begin{thebibliography}{99}
\bibitem{GORDON}
  W. Gordon, {\em Zeits. f. Physik\/} {\bf 48} (1928) 180.

\bibitem{BornW3}
  M. Born and E. Wolf, {\em Principles of Optics\/}, 3rd ed. (Pergamon,
  New York,  1965) Chapts. 3 and 4.

\bibitem{Brekhovskikh80}
 L. M. Brekhovskikh, {\em Waves in Layered Media\/}, 2nd ed.
   (Academic, New York, 1980).

\bibitem{KravOrlo80}
 Yu. A. Kravtsov and Yu. I. Orlov,  {\em Sov. Phys. Usp.\/} {\bf 23}
  (1980) 750.

\bibitem{Kravtsov88}
  Yu. A. Kravtsov, Rays and caustics as physical objects,
   in E. Wolf ed., {\em Progress in Optics\/}
   (North-Holland, Amsterdam, 1988) Vol. XXVI  229.

\bibitem{Tsong}
 T. T. Tsong,  {\em Atom-probe Field Ion Microscope\/},
 (Cambridge U. P., New York, 1990).

\bibitem{Ganz}
E. Ganz, S. K. Theiss, Ing-Shouh Hwang, and J. Golovchenko,
 {\em Phys. Rev. Lett.\/} {\bf 68} (1992) 1567.

\bibitem{Eigler}
 D. M. Eigler, C. P. Lutz and W. E. Rudge, {\em Nature\/}
 {\bf 352} (1991) 600.

\bibitem{Keller}
 J. B. Keller, {\em J. Opt. Soc. Am.\/} {\bf 52} (1962) 116.

\bibitem{KravOrlo93}
 Yu. A. Kravtsov and Yu. I. Orlov, {\em Caustics, Catastrophes and
   Wave Fields\/} (Springer, Berlin, 1993).

\bibitem{Nishiyama}
 Y. Nishiyama,  {\em J. Opt. Soc. Am. A \/} {\bf 12} (1995) 1390.

\bibitem{Goldstein}
 H. Goldstein, {\em Classical Mechanics\/} (Addison-Wesley, Reading, 1950),
  Chap. 9.

\bibitem{YN6365}
Preliminary attempts had been done:\,
Y. Nishiyama, {\em Prog. Theor. Phys. \/}{\bf 30} (1963) 657; {\bf 34},
 (1965) 299, 473.

\bibitem{Bohm}
 D. Bohm,  {\em Phys. Rev.\/} {\bf 85} (1952) 166; \
D. Bohm and B. J. Hiley, {\em Phys. Rep. \/} {\bf 144} (1987) 323. 

\bibitem{Holland}
P R Holland, {\em The quantum theory of motion \/} (Cambridge U. P.,
 Cambridge, 1993).

\bibitem{Bohm85}
D. Bohm, Hidden variables and the implicate order, in  B. J. Hiley
and F. D. Peat, eds., {\em Quantum implications, Essays in honour of
 David Bohm\/} (Routledge and Kegan Paul, London, 1987) 33.

\bibitem{KelRubi}
 J. B. Keller and S. I. Rubinow, {\em Ann. Physics\/}
  {\bf 9} (1960) 24.

\bibitem{AbraSteg}
 M. Abramowitz and  I. A. Stegun, {\em Handbook of Mathematical
  Functions\/} (Dover, New York, 1965), Chap.13.

\end{thebibliography}
\end{document}